# Leptoquark Pair Production at $ep$ Colliders

Johannes Blümlein[1], Edward Boos[1,2], and Alexander Pukhov[2]

[1]*DESY – Institut für Hochenergiephysik Zeuthen,
Platanenallee 6, D–15735 Zeuthen, Germany*
[2]*Institute of Nuclear Physics, Moscow State University,
RU–119899 Moscow, Russia*

## Abstract

The pair production cross section for scalar and vector leptoquarks at $ep$ colliders is calculated for the case of photon–gluon fusion. In a model independent analysis we consider the most general $C$ and $P$ conserving couplings of gluons and photons to both scalar and vector leptoquarks described by an effective low–energy Lagangian which obeys $U(1)_{em} \times SU(3)_c$ invariance. Numerical predictions are given for the kinematical regime at HERA and LEP $\otimes$ LHC.

# 1 Introduction

Many theories beyond the Standard Model try to unify the observed quark and lepton degrees of freedom on a more fundamental level [1]. As a consequence, new bosons, the leptoquarks, are contained in these models. For a long time $ep$ colliders have been considered as ideal facilities to search for leptoquarks through $e^{\pm}q(\overline{q})$ fusion [2, 3], since their signal emerges as a narrow peak in the deep inelastic differential $e^{\pm}p$ scattering cross sections $d^2\sigma^{e^{\pm}p}/dxdQ^2$. Searching for these peaks, the first experimental limits from collider experiments on the mass and the $l^{\pm}q(\overline{q})$ couplings, $\lambda_{lq}$, of leptoquarks to the fermions of the first generation have been given by ZEUS and H1 (cf. [4]) at HERA recently. The couplings $\lambda_{lq}$ are *not* predicted by theory and it is not excluded that $\lambda_{lq}/e \ll 1$. In fact, a recent re–analysis of different measurements with respect to leptoquark contributions [5] constrains $\lambda_{lq}/e \lesssim 0.07...0.27$, depending on the type of scalar and vector leptoquarks. For $\lambda_{lq}/e \ll 1$ both the production cross sections for $e^{\pm}q(\overline{q})$ fusion and $e^{\pm}g$ fusion [6] are rather small.

Leptoquark pair production via photon–gluon fusion depends on the gauge boson couplings to leptoquarks *only*. Thus, a dedicated search for leptoquarks is possible also in the range of small fermion couplings. In the case of scalar leptoquarks all couplings are known completely [1]. For vector leptoquarks the situation is more complex and depends on the specific nature of the low–energy leptoquark states emerging after symmetry breaking in a unified theory or in some scenario of compositeness. To keep the analysis as model independent as possible we assume the most general Lorentz structure for photon and gluon–leptoquark couplings which respect $C$ and $P$ conservation [8]. The cases of a minimal vector boson coupling (cf. [9]) and a Yang–Mills type coupling are contained in this description as well as the 'anomalous' couplings $\kappa_{A,G}$ and $\lambda_{A,G}$. These parameters determine the production cross sections in addition to the known electromagnetic and color couplings.

In this letter we derive the pair production cross sections for vector leptoquarks based on photon–gluon fusion (section 2). The expectations to produce the different types of scalar and vector leptoquarks at HERA and possible future experiments at LEP $\times$ LHC are discussed in section 3. An appendix summarizes the functions which describe the differential and integrated cross section of vector leptoquark pair production.

# 2 Production Cross Sections

We will calculate the contributions to leptoquark pair production due to photon–gluon fusion and consider the direct terms only. The resolved photon contributions are dealt with in a separate paper [10] [2]. The Feynman diagrams which determine both the scalar and vector leptoquark pair production cross sections are shown in figure 1. Note, that contrary to the case of $e^+e^-$ annihilation [9] Yukawa–type fermion couplings do not contribute. The interaction of the different leptoquarks species with photons and gluons is described by the effective Lagrangian in equ. (1) which is constructed to be invariant under $U(1)_{em} \times SU(3)_c$ gauge transformations.

$$\mathcal{L} = \mathcal{L}_s + \mathcal{L}_v \qquad (1)$$

---

[1] The structure of the scattering cross section for this case has been known from scalar electrodynamics for a very long time [7]. We include a brief discussion of scalars only for the purpose of a systematic comparison with respect to the classification of leptoquark states [3] and as a numerical update.

[2] At high virtualities, $Q^2$, of the intermediate boson terms due to $\gamma$–$Z$ interference and $Z$–exchange also become relevant. Furthermore, pairs of different type leptoquarks (cf. [9]) can be produced via $W^{\pm}g$ fusion in this kinematical range.

with
$$\mathcal{L}_s = \sum_{scalars} \left[ (D^\mu \Phi)^\dagger (D_\mu \Phi) - M_s^2 \Phi^\dagger \Phi \right] \tag{2}$$

and [3]

$$\begin{aligned}
\mathcal{L}_v &= \sum_{vectors} \left\{ -\frac{1}{2} G^\dagger_{\mu\nu} G^{\mu\nu} + M_v^2 \Phi^\dagger_\mu \Phi^\mu - ie \left[ (1 - \kappa_A) \Phi^\dagger_\mu \Phi_\nu F^{\mu\nu} + \frac{\lambda_A}{M_v^2} G^\dagger_{\sigma\mu} G^\mu_\nu F^{\nu\sigma} \right] \right. \\
&\quad - \left. ig_s \left[ (1 - \kappa_G) \Phi^\dagger_\mu \frac{\lambda^a}{2} \Phi_\nu \mathcal{G}^{\mu\nu}_a + \frac{\lambda_G}{M_v^2} G^\dagger_{\sigma\mu} \frac{\lambda^a}{2} G^\mu_\nu \mathcal{G}^{\nu\sigma}_a \right] \right\}.
\end{aligned} \tag{3}$$

Here $e$ and $g_s$ denote the electromagnetic and strong coupling constant, and $\kappa_{A,G}$ and $\lambda_{A,G}$ are the anomalous couplings. The field strength tensors of the photon–, gluon–, and vector leptoquark fields are

$$\begin{aligned}
F_{\mu\nu} &= \partial_\mu A_\nu - \partial_\nu A_\mu, \\
\mathcal{G}^a_{\mu\nu} &= \partial_\mu \mathcal{A}^a_\nu - \partial_\nu \mathcal{A}^a_\mu + if^{abc} \mathcal{A}_{\mu b} \mathcal{A}_{\nu c}, \\
G_{\mu\nu} &= D_\mu \Phi_\nu - D_\nu \Phi_\mu
\end{aligned} \tag{4}$$

with the covariant derivative given as

$$D_\mu = \partial_\mu - ieQ^\gamma A_\mu - ig_s \frac{\lambda^a}{2} \mathcal{A}^a_\mu. \tag{5}$$

The parameters $\kappa_{A,G}$ and $\lambda_{A,G}$ are assumed to be real. They are related to the anomalous 'magnetic' moment $\mu_\Phi$ and 'electric' quadrupole moment $q_\Phi$ of the leptoquarks in the electromagnetic and color fields

$$\begin{aligned}
\mu_{\Phi,\alpha} &= \frac{g_\alpha}{2M_\Phi} (2 - \kappa_\alpha + \lambda_\alpha) \\
q_{\Phi,\alpha} &= -\frac{g_\alpha}{M_\Phi^2} (1 - \kappa_\alpha - \lambda_\alpha)
\end{aligned} \tag{6}$$

where $g_\alpha = e$ or $g_s$ and $\alpha = A$ or $G$ [4].

In particular we assume that these quantities are all independent since we wish to keep the analysis as model independent as possible.

We consider the leptoquarks which have been classified in [3, 9]. They are color triplets or anti–triplets, and the magnitude of their electric charges $|Q_\Phi|$ can take the values 5/3, 4/3, 2/3, or 1/3. The cross sections are calculated using the (improved) Weizsäcker–Williams approximation (WWA) [12].

The integrated pair production cross sections read

$$\sigma_{s,v}(S, M_\Phi^2) = \int_{y_{min}}^{y_{max}} dy \int_{x_{min}}^{x_{max}} dx \int_{-1}^{1} d\cos\theta \; \phi_{\gamma/e}(y) G_{g/p}(x, \mu^2) \frac{d\hat{\sigma}_{s,v}}{d\cos\theta} \theta(\hat{s} - 4M_\Phi^2). \tag{7}$$

Here $d\hat{\sigma}_{s,v}/d\cos\theta$ denotes the differential cross section in the photon–gluon center–of–momentum system (cms), $G_{g/p}(x,\mu^2)$ is the gluon distribution at the factorization mass $\mu$, $S = 4E_e E_p$, $\hat{s} = xyS$, $x$ is the longitudinal momentum fraction of the proton carried by the gluon, and $M_\Phi$ denotes the mass of the leptoquarks. The photon distribution is described in WWA by

$$\phi_{\gamma/e}(y) = \frac{\alpha}{2\pi}\left[2m_e^2 y\left(\frac{1}{Q_{max}^2} - \frac{1}{Q_{min}^2}\right) + \frac{1+(1-y)^2}{y}\log\frac{Q_{max}^2}{Q_{min}^2}\right]. \qquad (8)$$

---
[3] In compositeness scenarios one might wish to relate the quantities $\kappa_{A,G}$ and $\lambda_{A,G}$ to the compositeness scale $\Lambda$. This can be achieved e.g. rescaling these quantities by factor $M_v^2/\Lambda^2$.

[4] Note that the convention for the $\kappa_A$ and $\lambda_A$ used here translates into that of [11] by substituting $\kappa_A = 1 - \kappa_\gamma$, $\lambda_A = \lambda_\gamma$.

The kinematical boundaries in (7) and (8) are:

$$Q_{min}^2 = \frac{m_e^2 y^2}{1-y} \qquad Q_{max}^2 = yS - 4M_\Phi^2 - 4M_\Phi m_p$$

$$x_{min} = \frac{4M_\Phi^2}{yS} \qquad x_{max} = 1$$

$$y_{min,max} = \frac{S + \widetilde{W}^2 \pm \sqrt{(S - \widetilde{W}^2)^2 - 4m_e^2 \widetilde{W}^2}}{2(S + m_e^2)} \qquad (9)$$

where $\widetilde{W}^2 = (2M_\Phi + m_p)^2 - m_p^2$, $m_e$ and $m_p$ are the electron and proton mass, respectively, $y = P.q/P.l_e$, with $q = l_e - l'_e$, and $P, l_e, l'_e$ the four momenta of the proton, the incoming and outgoing electron.

## 2.1 Scalar Leptoquarks

The differential and integrated production cross sections in the $\gamma$–$g$ cms are[5]

$$\frac{d\hat{\sigma}_s}{d\cos\theta} = \frac{\pi\alpha\alpha_s(\mu^2)}{2\hat{s}}Q_\Phi^2 \beta\left\{1 - \frac{2(1-\beta^2)}{1-\beta^2\cos^2\theta} + \frac{2(1-\beta^2)^2}{(1-\beta^2\cos^2\theta)^2}\right\} \qquad (10)$$

and

$$\hat{\sigma}_s(\hat{s}, \beta) = \frac{\pi\alpha\alpha_s(\mu^2)}{2\hat{s}}Q_\Phi^2 \left\{2(2-\beta^2)\beta - (1-\beta^4)\log\left|\frac{1+\beta}{1-\beta}\right|\right\}. \qquad (11)$$

Here, $\beta = \sqrt{1 - 4M_\Phi^2/\hat{s}}$ and $\alpha_s$ denotes the strong coupling constant. The production cross section varies $\propto Q_\Phi^2$. For the leptoquark states classified in [3] one obtains e.g. $\sigma_s(R_2^{5/3}) = 25\,\sigma_s(S_3^{1/3})$, etc.[6].

## 2.2 Vector Leptoquarks

The corresponding cross sections for vector leptoquarks can be represented in terms of the individual $A \leftrightarrow G$ symmetric combinations of $\kappa_{A,G}$ and $\lambda_{A,G}$ at tree level. The differential cross section in the $\gamma$–$g$ cms is

$$\frac{d\hat{\sigma}_v}{d\cos\theta} = \frac{\pi\alpha\alpha_s(\mu^2)}{2\hat{s}}Q_\Phi^2 \sum_{j=0}^{20}\chi_j(\kappa_{A,G}, \lambda_{A,G})\frac{F_j(\hat{s}, \beta, \cos\theta)}{(1-\beta^2\cos^2\theta)^2} \qquad (12)$$

---

[5]The form of the cross section (10,11) has been derived in [7] and was used mutually in the literature for various processes [13] with different couplings and group theoretical factors.

[6]In the case of leptoquark pair production through $\gamma\gamma$ fusion [14], the ratio of the production cross sections varies even by factors up to 625.

with

$$
\begin{aligned}
\sum_{j=0}^{20} \chi_j(\kappa_{A,G}, \lambda_{A,G}) F_j =\ & F_0 & +\ & (\kappa_A + \kappa_G) F_1 \\
& + (\kappa_A^2 + \kappa_G^2) F_2 & +\ & \kappa_A \kappa_G F_3 \\
& + \kappa_A \kappa_G (\kappa_A + \kappa_G) F_4 & +\ & \kappa_A^2 \kappa_G^2 F_5 \\
& + (\lambda_A + \lambda_G) F_6 & +\ & (\lambda_A^2 + \lambda_G^2) F_7 \\
& + \lambda_A \lambda_G F_8 & +\ & \lambda_A \lambda_G (\lambda_A + \lambda_G) F_9 \\
& + \lambda_A^2 \lambda_G^2 F_{10} & +\ & (\kappa_A \lambda_A + \kappa_G \lambda_G) F_{11} \\
& + (\kappa_A \lambda_G + \kappa_G \lambda_A) F_{12} & +\ & \lambda_A \lambda_G (\kappa_A + \kappa_G) F_{13} \\
& + (\kappa_A \lambda_G^2 + \kappa_G \lambda_A^2) F_{14} & +\ & \lambda_A \lambda_G (\kappa_A \lambda_G + \kappa_G \lambda_A) F_{15} \\
& + (\kappa_A^2 \lambda_G + \kappa_G^2 \lambda_A) F_{16} & +\ & (\kappa_A^2 \lambda_G^2 + \kappa_G^2 \lambda_A^2) F_{17} \\
& + \kappa_A \kappa_G (\lambda_A + \lambda_G) F_{18} & +\ & \kappa_A \kappa_G \lambda_A \lambda_G F_{19} \\
& + \kappa_A \kappa_G (\kappa_A \lambda_G + \kappa_G \lambda_A) F_{20}
\end{aligned}
$$

and the integrated cross section reads

$$\hat{\sigma}_v = \frac{\pi \alpha \alpha_s(\mu^2)}{M_v^2} Q_\Phi^2 \sum_{j=0}^{20} \chi_j(\kappa_{A,G}, \lambda_{A,G}) \widetilde{F}_j(\hat{s}, \beta) \qquad (13)$$

where

$$\widetilde{F}_j = \frac{M_v^2}{\hat{s}} \int_0^\beta d\xi \frac{F_j(\xi = \beta \cos\theta)}{(1-\xi^2)^2}. \qquad (14)$$

The functions $F_j(\hat{s}, \beta, \cos\theta)$ and $\widetilde{F}_j(\hat{s}, \beta)$ are obtained in a lengthly but straightforward calculation which has been performed using the package CompHEP [15]. They are given in eqs. (17,18) in the appendix.

For the case $\kappa_A = \kappa_G$ and $\lambda_A = \lambda_G$ the equations (17) agree with results obtained in [11] for the case of $W$-boson pair production in $\gamma$–$\gamma$ fusion. Other more specific results derived earlier for particular choices of $\kappa$ and $\lambda$ [16] are described by (12). In our notation the case $\kappa_{A,G} = \lambda_{A,G} \equiv 0$ corresponds to Yang–Mills type couplings of photons and gluons to vector leptoquarks, while $\kappa_{A,G} = 1, \lambda_{A,G} = 0$ describes the case of 'minimal' vector boson couplings [17]. Most of the above terms contain contributions $\propto (\hat{s}/M_\Phi^2)^n$ which are of $\mathcal{O}(1)$ in the threshold range. These unitarity–violating terms (for $\hat{s} \gg M_v^2$) are absent in some of the functions $\widetilde{F}_j$, particularly for the contributions which are at most *linear* in $\kappa_{A,G}$ and $\lambda_{A,G}$.

One may ask whether apart from the Yang–Mills case another combination of these couplings exists which preserves tree–level unitarity. Since only in $\widetilde{F}_{10}$ a term $\propto (\hat{s}/M_\Phi^2)^3$ appears either $\lambda_A$ or $\lambda_G$ must vanish to preserve unitarity. Furthermore, the terms $\propto (\hat{s}/M_\Phi^2)^2$ cancel only for

$$\lambda_{A(G)}^2 \left[1 + (1 - \kappa_{G(A)}^2)^2\right] = 0 \qquad (15)$$

Both $\lambda_A$ and $\lambda_G$ have to vanish to obtain real solutions. The terms $\propto (\hat{s}/M_\Phi^2)$ cancel if $\kappa_A$ and $\kappa_G$ obey the relation

$$(\kappa_A + \kappa_G - \kappa_A \kappa_G)^2 + \frac{2}{3} \kappa_A^2 \kappa_G^2 = 0. \qquad (16)$$

Thus, for any non vanishing values of $\kappa_A, \kappa_G, \lambda_A, \lambda_G$ tree–level unitarity is not preserved by $\sigma_v$. At high energies (i.e. $\hat{s} \gg 4M_v^2$) the effective low energy Lagrangian (1) is no longer valid since terms which decouple at low energies become relevant. Instead one has to consider the full gauge theory from which (1) was obtained.

# 3   Numerical Results

In the subsequent numerical calculations we used the CTEQ2 (LO) parametrization to describe the gluon distribution [18]. Other recent parametrizations [19] yield similar numerical results. The factorization mass $\mu$ and the scale of $\alpha_s$ were choosen to be $\sqrt{s}$. The uncertainty of the cross section calculation due to the use of the improved Weizsäcker–Williams approximation [12] was estimated to be of $\mathcal{O}(6\%)$ both for the case of HERA and LEP $\times$ LHC due to the choice of the kinematical bounds (9a).

In figure 2a the integrated cross section for $ep$ scattering at HERA is shown for scalar leptoquarks with $|Q_\Phi| = 1/3$ and $5/3$, respectively. Production cross sections $\sigma_s^{tot} \gtrsim 0.1 pb$ – corresponding to 10 events at an integrated luminosity of $\mathcal{L} = 100 pb^{-1}$ – are obtained for the states with $|Q_\Phi| = 2/3$ to $|Q_\Phi| = 5/3$ for $M_\Phi <$ 55 to 65 GeV. For $|Q_\Phi| = 1/3$ the production cross section is too small for $M_\Phi \gtrsim 45$ GeV, a bound set previously by the LEP experiments [20]. At LEP $\times$ LHC the search limits on a rate of 10 events extend from $M_\Phi = 140$ GeV for $|Q_\Phi| = 1/3$ to $M_\Phi = 220$ GeV for $|Q_\Phi| = 5/3$ assuming $\mathcal{L} = 1 fb^{-1}$ as shown in figure 2b.

Limits on the allowed mass range of scalar leptoquarks have been derived by different experiments. Bounds which are independent of the leptoquark–fermion couplings were given by the LEP experiments as $M_\Phi > 44.4$ GeV at 95 % CL [20, 21][7] for almost all scalars classified in [3] and all three generations. Other limits have been found by the UA2 [23], CDF and D0 [24] experiments for the 1st generation scalar leptoquarks, excluding the mass ranges between 44 and 132 GeV depending on the branching ratios $Br(\Phi_s \to eq)$. Independently of the yet unknown fermion couplings $\lambda_{L,R}$, the states $\widetilde{S}_1, S_3^{4/3}, R_2^{5/3}$ and $\widetilde{R}_2^{2/3}$ are excluded for masses $M_\Phi < 132$ GeV and the state $S_1$ for $M_\Phi < 86$ GeV. For general values of the fermion couplings, the states $S_3^{-2/3}, \widetilde{R}_2^{-1/3}, R_2^{2/3}$ and $S_3^{1/3}$ are constrained by the LEP bound only. Unlike these partial results a systematic search for scalar leptoquarks of *all* generations above the LEP limit is still needed in the kinematically accessible range at HERA, $M_\Phi \lesssim 63$ GeV. Among the different states [3] the search for $S_3^{-2/3}$ and $\widetilde{R}_2^{2/3}$ (with $\lambda_R = 0$) (cf. [9]) is particularly difficult, since these leptoquarks decay only into a neutrino–quark pair, a signature with a large QCD background.

Figure 3a shows the integrated cross section for vector leptoquark pair production at HERA for different choices of $\kappa_{A,G}$ and $\lambda_{A,G}$. In a model independent analysis the complete dependences on these four parameters must be explored. From the examples shown in figure 3a it is evident, that severe constraints on the parameter space of the anomalous couplings can be obtained at HERA. Complementary to searches at $p\bar{p}$ colliders the photon couplings $\kappa_A$ and $\lambda_A$ can be probed besides of $\kappa_G$ and $\lambda_G$ at $ep$ colliders. Due to accidental cancellations between the different contributions, $\widetilde{F}_j$, for specific values of $\kappa_{A,G}, \lambda_{A,G}$ even smaller cross sections than that of the minimal vector coupling (M.C.) can be obtained. The cross section $\sigma_{M.C.}^{tot}$ for the case of HERA amounts to 0.2 $pb$ only at the LEP bound[8]. For minimal vector coupling, the search limits at a rate of 10 events extend to 52 GeV ($|Q_\Phi| = 1/3$) and to 74 GeV ($|Q_\Phi| = 5/3$). Figure 3b shows the mass dependence of the integrated cross sections for vector leptoquark pair production at LEP $\times$ LHC with $|Q_\Phi| = 1/3$. The search limits on a rate of 10 events for leptoquarks coupling to both photons and gluons with a minimal vector coupling extend from $M_\Phi = 185$ GeV to

---

[7]In [21] also limits on 1st and 2nd generation scalar leptoquarks are derived from a search in $e^+e^- \to \overline{S}S^* \hookrightarrow lq$ [22]. However, these bounds depend on assumptions made on $\lambda_{lq}$.

[8]Neither the LEP experiments nor searches at proton colliders have investigated vector leptoquarks so far allowing for general vector boson–gauge boson couplings (3). Since there is a symmetry in the decay pattern of scalar and vector leptoquarks (cf. [9], table 3), one finds from the cross sections calculated in [9] that the exclusion limit found for scalars at LEP holds for vectors as well.

270 GeV for $|Q_\Phi| = 1/3$ to $|Q_\Phi| = 5/3$ vector leptoquarks.

Note that the above bounds have been calculated on the basis of the direct contributions due to photon–gluon fusion only. The resolved photon contributions [10] will allow to extend the mass ranges correspondingly.

In summary we have shown that in a search for both scalar and vector leptoquark pair production at $ep$ colliders yet open mass ranges can be explored independently of the size of the leptoquark–fermion couplings and fermion generation to which the leptoquarks are associated. For vector leptoquarks constraints on both the anomalous $\kappa_A, \lambda_A$ and $\kappa_G, \lambda_G$ can be derived.

**Acknowledgement.** We are grateful to Slava Ilyin and Sergey Shichanin for discussions. We would like to thank Günter Wolf and Peter Zerwas for conversations, and James Botts for reading the manuscript. E.B. would like to thank DESY–Zeuthen for the warm hospitality extended to him.

## 4  Appendix

The functions $F_i(\hat{s}, \beta, \cos\theta)$ of (12) are:

$$F_0 = 19 - 6\beta^2 + 6\beta^4 + (16 - 6\beta^2)\beta^2 \cos^2\theta + 3\beta^4 \cos^4\theta$$

$$F_1 = -22 - 10\beta^2 \cos^2\theta$$

$$F_2 = 4 + \frac{\hat{s}}{M_\Phi^2} \frac{1 - \beta^4 \cos^4\theta}{2} + \frac{\hat{s}^2}{M_\Phi^4} \frac{(1 - \beta^2 \cos^2\theta)^2}{16}$$

$$F_3 = 28 + 4\beta^2 \cos^2\theta + \frac{\hat{s}}{M_\Phi^2}\beta^2 \cos^2\theta(1 - \beta^2 \cos^2\theta) + \frac{\hat{s}^2}{M_\Phi^4}\frac{(1 - \beta^2 \cos^2\theta)^2}{8}$$

$$F_4 = -5 + \beta^2 \cos^2\theta + \frac{\hat{s}}{M_\Phi^2}\frac{-3 + \beta^2 \cos^2\theta + 2\beta^4 \cos^4\theta}{4} - \frac{\hat{s}^2}{M_\Phi^4}\frac{(1 - \beta^2 \cos^2\theta)^2}{8}$$

$$F_5 = \frac{3 - \beta^2 \cos^2\theta}{4} + \frac{\hat{s}}{M_\Phi^2}\frac{5 - 4\beta^2 \cos^2\theta - \beta^4 \cos^4\theta}{16}$$
$$+ \frac{\hat{s}^2}{M_\Phi^4}\frac{13 - 25\beta^2 \cos^2\theta + 11\beta^4 \cos^4\theta + \beta^6 \cos^6\theta}{128}$$

$$F_6 = -4 + 4\beta^2 \cos^2\theta$$

$$F_7 = 4 + \frac{\hat{s}}{M_\Phi^2}\frac{-7 + 8\beta^2 \cos^2\theta - \beta^4 \cos^4\theta}{2} + \frac{\hat{s}^2}{M_\Phi^4}\frac{(1 - \beta^2 \cos^2\theta)^2}{2}$$
$$+ \frac{\hat{s}^3}{M_\Phi^6}\frac{(1 - \beta^2 \cos^2\theta)^3}{16}$$

$$F_8 = -\frac{\hat{s}}{M_\Phi^2}(1 - \beta^2 \cos^2\theta) + \frac{\hat{s}^2}{M_\Phi^4}\frac{11 - 13\beta^2 \cos^2\theta + \beta^4 \cos^4\theta + \beta^6 \cos^6\theta}{8}$$
$$- \frac{\hat{s}^2}{M_\Phi^4}\frac{(1 - \beta^2 \cos^2\theta)^3}{8}$$

$$F_9 = 1 - \beta^2 \cos^2\theta + \frac{\hat{s}}{M_\Phi^2}\frac{-3 + 4\beta^2 \cos^2\theta - \beta^4 \cos^4\theta}{2} + \frac{\hat{s}^2}{M_\Phi^4}(1 - \beta^2 \cos^2\theta)^2$$
$$+ \frac{\hat{s}^3}{M_\Phi^6}\frac{-3 + 7\beta^2 \cos^2\theta - 5\beta^4 \cos^4\theta + \beta^6 \cos^6\theta}{16}$$

$$F_{10} = \frac{3 - \beta^2 \cos^2\theta}{4} + \frac{\hat{s}}{M_\Phi^2} \frac{-19 + 20\beta^2 \cos^2\theta - \beta^4 \cos^4\theta}{16}$$

$$+ \frac{\hat{s}^2}{M_\Phi^4} \frac{141 - 249\beta^2 \cos^2\theta + 107\beta^4 \cos^4\theta + \beta^6 \cos^6\theta}{128}$$

$$+ \frac{\hat{s}^3}{M_\Phi^6} \frac{-53 + 119\beta^2 \cos^2\theta - 79\beta^4 \cos^4\theta + 13\beta^6 \cos^6\theta}{128}$$

$$+ \frac{\hat{s}^4}{M_\Phi^8} \frac{27 - 68\beta^2 \cos^2\theta + 58\beta^4 \cos^4\theta - 20\beta^6 \cos^6\theta + 3\beta^8 \cos^8\theta}{512}$$

$$F_{11} = -8 + \frac{\hat{s}}{M_\Phi^2}(3 - 4\beta^2 \cos^2\theta + \beta^4 \cos^4\theta)$$

$$F_{12} = \frac{\hat{s}}{M_\Phi^2}(2 - 3\beta^2 \cos^2\theta + 4\beta^4 \cos^4\theta)$$

$$F_{13} = -2(1 - \beta^2 \cos^2\theta) + \frac{\hat{s}}{M_\Phi^2} \frac{9 - 13\beta^2 \cos^2\theta + 4\beta^4 \cos^4\theta}{4}$$

$$- \frac{\hat{s}^2}{M_\Phi^4} \frac{2 - 3\beta^2 \cos^2\theta + \beta^4 \cos^4\theta}{2} + \frac{\hat{s}^3}{M_\Phi^6} \frac{(1 - \beta^2 \cos^2\theta)^3}{16}$$

$$F_{14} = -5 + \beta^2 \cos^2\theta + \frac{\hat{s}}{M_\Phi^2} \frac{7 - 8\beta^2 \cos^2\theta + \beta^4 \cos^4\theta}{2} - 3\frac{\hat{s}^2}{M_\Phi^4} \frac{(1 - \beta^2 \cos^2\theta)^2}{8}$$

$$- \frac{(1 - \beta^2 \cos^2\theta)^3}{16}$$

$$F_{15} = -\frac{3 - \beta^2 \cos^2\theta}{2} + \frac{\hat{s}}{M_\Phi^2} \frac{13 - 14\beta^2 \cos^2\theta + \beta^4 \cos^4\theta}{8}$$

$$- \frac{\hat{s}^2}{M_\Phi^4} \frac{41 - 81\beta^2 \cos^2\theta + 39\beta^4 \cos^4\theta + \beta^6 \cos^6\theta}{64}$$

$$+ \frac{\hat{s}^3}{M_\Phi^6} \frac{11 - 25\beta^2 \cos^2\theta + 17\beta^4 \cos^4\theta - 3\beta^6 \cos^6\theta}{128}$$

$$F_{16} = 1 - \beta^2 \cos^2\theta - \frac{\hat{s}}{M_\Phi^2} \frac{3 - 5\beta^2 \cos^2\theta + 2\beta^4 \cos^4\theta}{4}$$

$$F_{17} = \frac{3 - \beta^2 \cos^2\theta}{4} - \frac{\hat{s}}{M_\Phi^2} \frac{7 - 8\beta^2 \cos^2\theta + \beta^4 \cos^4\theta}{16}$$

$$- \frac{\hat{s}^2}{M_\Phi^4} \frac{3 - 7\beta^2 \cos^2\theta + 5\beta^4 \cos^4\theta - \beta^6 \cos^6\theta}{128} + \frac{\hat{s}^3}{M_\Phi^6} \frac{(1 - \beta^2 \cos^2\theta)^3}{32}$$

$$F_{18} = 2(5 - \beta^2 \cos^2\theta) - \frac{\hat{s}}{M_\Phi^2} \frac{11 - 15\beta^2 \cos^2\theta + 4\beta^4 \cos^4\theta}{4}$$

$$- \frac{\hat{s}^2}{M_\Phi^4} \frac{(1 - \beta^2 \cos^2\theta)^2}{4}$$

$$F_{19} = 3 - \beta^2 \cos^2\theta - \frac{\hat{s}}{M_\Phi^2} \frac{7 - 8\beta^2 \cos^2\theta + \beta^4 \cos^4\theta}{4}$$

$$+ \frac{\hat{s}^2}{M_\Phi^4} \frac{11 - 13\beta^2 \cos^2\theta + \beta^4 \cos^4\theta + \beta^6 \cos^6\theta}{32}$$

$$+ \frac{\hat{s}^3}{M_\Phi^6} \frac{5 - 7\beta^2 \cos^2\theta - \beta^4 \cos^4\theta + 3\beta^6 \cos^6\theta}{128}$$

$$F_{20} = -\frac{3 - \beta^2 \cos^2 \theta}{2} + \frac{\hat{s}}{M_\Phi^2} \frac{(1 - \beta^2 \cos^2 \theta)^2}{8}$$
$$+ \frac{\hat{s}^2}{M_\Phi^4} \frac{11 - 23\beta^2 \cos^2 \theta + 13\beta^4 \cos^4 \theta - \beta^6 \cos^6 \theta}{64} \tag{17}$$

The functions $\widetilde{F}_i(\hat{s}, \beta)$, which describe the different contributions to the integrated cross section (13), are:

$$\widetilde{F}_0 = \beta \left( \frac{11}{2} - \frac{9}{4}\beta^2 + \frac{3}{4}\beta^4 \right) - \frac{3}{8} \left( 1 - \beta^2 - \beta^4 + \beta^6 \right) \ln \left| \frac{1+\beta}{1-\beta} \right|$$

$$\widetilde{F}_1 = -4\beta - \frac{3}{4} \left( 1 - \beta^2 \right) \log \left| \frac{1+\beta}{1-\beta} \right|$$

$$\widetilde{F}_2 = \frac{1}{16} \beta \frac{\hat{s}}{M_\Phi^2} + \frac{3 - \beta^2}{4} \log \left| \frac{1+\beta}{1-\beta} \right|$$

$$\widetilde{F}_3 = 3\beta + \frac{1}{8} \beta \frac{\hat{s}}{M_\Phi^2} + \left( 2 - \frac{3}{2}\beta^2 \right) \log \left| \frac{1+\beta}{1-\beta} \right|$$

$$\widetilde{F}_4 = -\frac{1}{8} \beta \frac{\hat{s}}{M_\Phi^2} + \left( -1 + \frac{3}{8}\beta^2 \right) \log \left| \frac{1+\beta}{1-\beta} \right|$$

$$\widetilde{F}_5 = -\frac{1}{96}\beta + \frac{5}{48} \beta \frac{\hat{s}}{M_\Phi^2} + \frac{4 - \beta^2}{16} \log \left| \frac{1+\beta}{1-\beta} \right|$$

$$\widetilde{F}_6 = -\frac{1}{2} \left( 1 - \beta^2 \right) \log \left| \frac{1+\beta}{1-\beta} \right|$$

$$\widetilde{F}_7 = \frac{7}{12} \beta \frac{\hat{s}}{M_\Phi^2} + \frac{1}{24} \beta \frac{\hat{s}^2}{M_\Phi^4} - \frac{5 + \beta^2}{4} \log \left| \frac{1+\beta}{1-\beta} \right|$$

$$\widetilde{F}_8 = -\frac{1}{6}\beta + \frac{1}{4} \beta \frac{\hat{s}}{M_\Phi^2} - \frac{1}{12} \beta \frac{\hat{s}^2}{M_\Phi^4} + \left( -\frac{1}{2} + \frac{1}{2} \frac{\hat{s}}{M_\Phi^2} \right) \log \left| \frac{1+\beta}{1-\beta} \right|$$

$$\widetilde{F}_9 = -\frac{1}{2}\beta + \frac{11}{12} \beta \frac{\hat{s}}{M_\Phi^2} - \frac{1}{6} \beta \frac{\hat{s}^2}{M_\Phi^4} - \frac{3 + \beta^2}{8} \log \left| \frac{1+\beta}{1-\beta} \right|$$

$$\widetilde{F}_{10} = -\frac{1}{96}\beta + \frac{59}{80} \beta \frac{\hat{s}}{M_\Phi^2} - \frac{113}{320} \beta \frac{\hat{s}^2}{M_\Phi^4} + \frac{43}{960} \beta \frac{\hat{s}^3}{M_\Phi^6} + \left( -\frac{1}{2} - \frac{1}{16}\beta^2 + \frac{1}{8} \frac{\hat{s}}{M_\Phi^2} \right) \log \left| \frac{1+\beta}{1-\beta} \right|$$

$$\widetilde{F}_{11} = \frac{1}{2} \left( 1 + \beta^2 \right) \log \left| \frac{1+\beta}{1-\beta} \right|$$

$$\widetilde{F}_{12} = \beta + \frac{1}{2} \log \left| \frac{1+\beta}{1-\beta} \right|$$

$$\widetilde{F}_{13} = \beta - \frac{5}{12} \beta \frac{\hat{s}}{M_\Phi^2} + \frac{1}{24} \beta \frac{\hat{s}^2}{M_\Phi^4} + \left[ -\frac{1}{4} \frac{\hat{s}}{M_\Phi^2} + \left( \frac{3}{8} + \frac{1}{4}\beta^2 \right) \right] \log \left| \frac{1+\beta}{1-\beta} \right|$$

$$\widetilde{F}_{14} = -\frac{11}{24} \beta \frac{\hat{s}}{M_\Phi^2} - \frac{1}{24} \beta \frac{\hat{s}^2}{M_\Phi^4} + \frac{9 + 3\beta^2}{8} \log \left| \frac{1+\beta}{1-\beta} \right|$$

$$\widetilde{F}_{15} = \frac{1}{48}\beta - \frac{59}{96} \beta \frac{\hat{s}}{M_\Phi^2} + \frac{5}{64} \beta \frac{\hat{s}^2}{M_\Phi^4} + \frac{5 + \beta^2}{8} \log \left| \frac{1+\beta}{1-\beta} \right|$$

$$\widetilde{F}_{16} = -\frac{1}{2}\beta - \frac{1}{8}\beta^2 \log \left| \frac{1+\beta}{1-\beta} \right|$$

$$\begin{aligned}
\widetilde{F}_{17} &= -\frac{1}{96}\beta + \frac{1}{48}\beta\frac{\hat{s}}{M_\Phi^2} + \frac{1}{48}\beta\frac{\hat{s}^2}{M_\Phi^4} - \frac{2+\beta^2}{16}\log\left|\frac{1+\beta}{1-\beta}\right| \\
\widetilde{F}_{18} &= -\frac{1}{4}\beta\frac{\hat{s}}{M_\Phi^2} - \frac{1-6\beta^2}{8}\log\left|\frac{1+\beta}{1-\beta}\right| \\
\widetilde{F}_{19} &= -\frac{1}{24}\beta + \frac{7}{96}\beta\frac{\hat{s}}{M_\Phi^2} + \frac{3}{64}\beta\frac{\hat{s}^2}{M_\Phi^4} + \left[\frac{1}{8}\frac{\hat{s}}{M_\Phi^2} - \frac{2+\beta^2}{4}\right]\log\left|\frac{1+\beta}{1-\beta}\right| \\
\widetilde{F}_{20} &= \frac{1}{48}\beta + \frac{1}{6}\beta\frac{\hat{s}}{M_\Phi^2} - \frac{1}{8}\left(1-\beta^2\right)\log\left|\frac{1+\beta}{1-\beta}\right|
\end{aligned} \qquad (18)$$

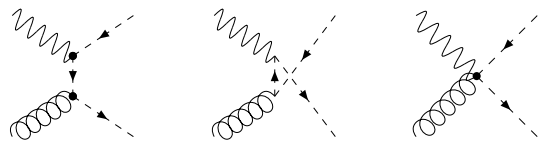

Figure 1: Diagrams describing leptoquark pair production via $\gamma$–$g$ fusion

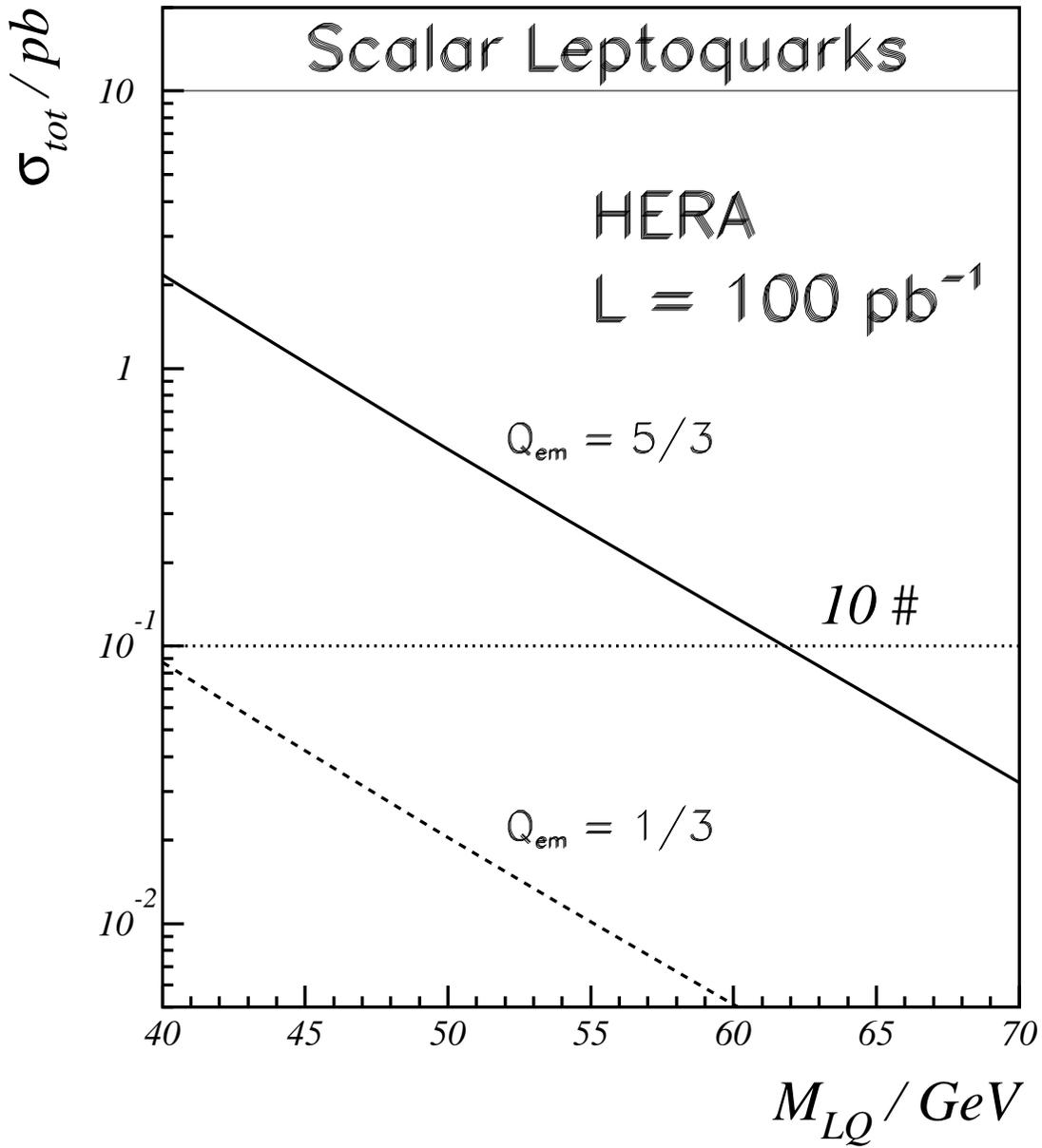

Figure 2a: Integrated cross sections for scalar leptoquark pair production, $\sqrt{S} = 314$ GeV.

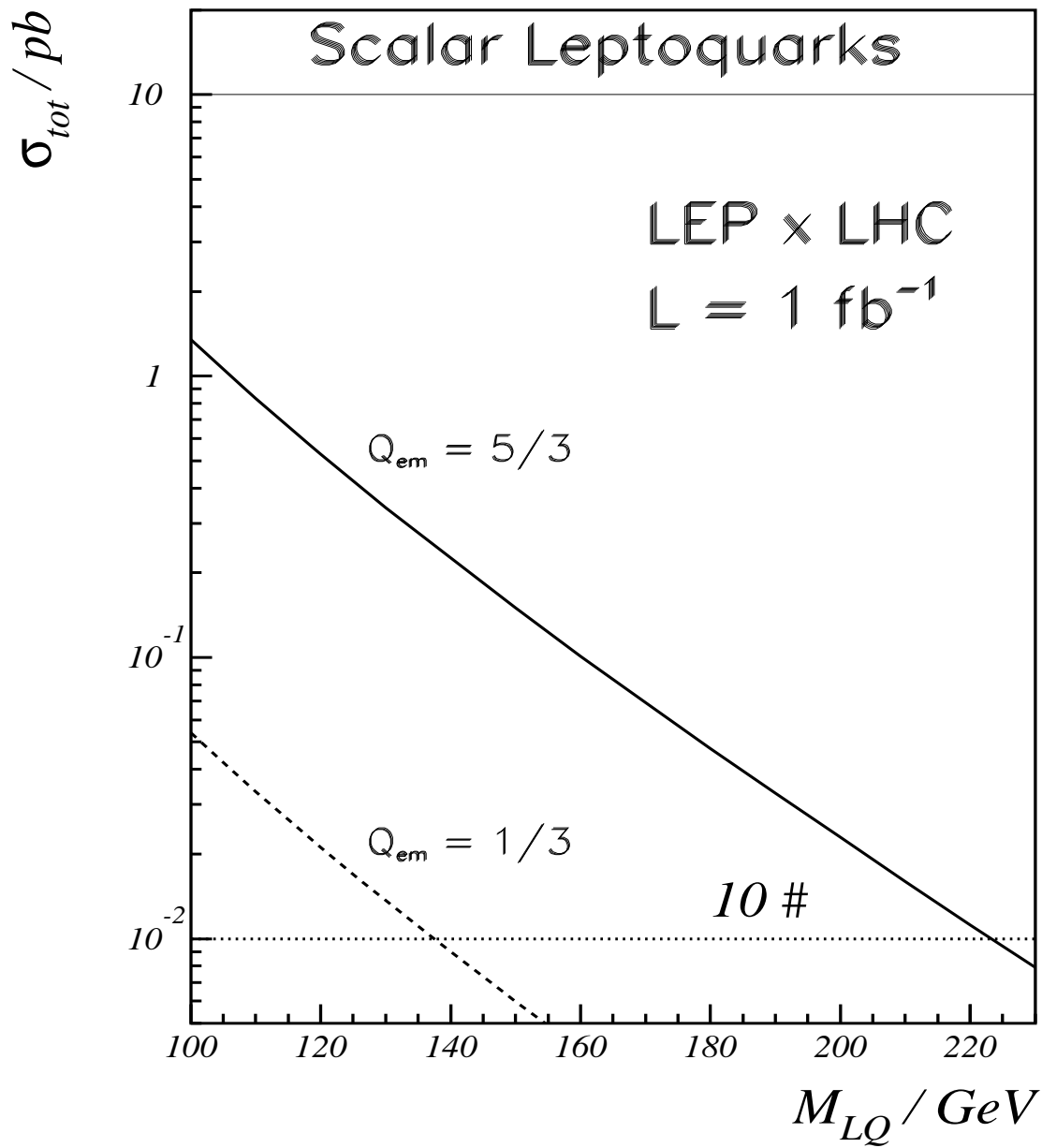

Figure 2b: Integrated cross sections for scalar leptoquark pair production, $\sqrt{S} = 1260$ GeV.

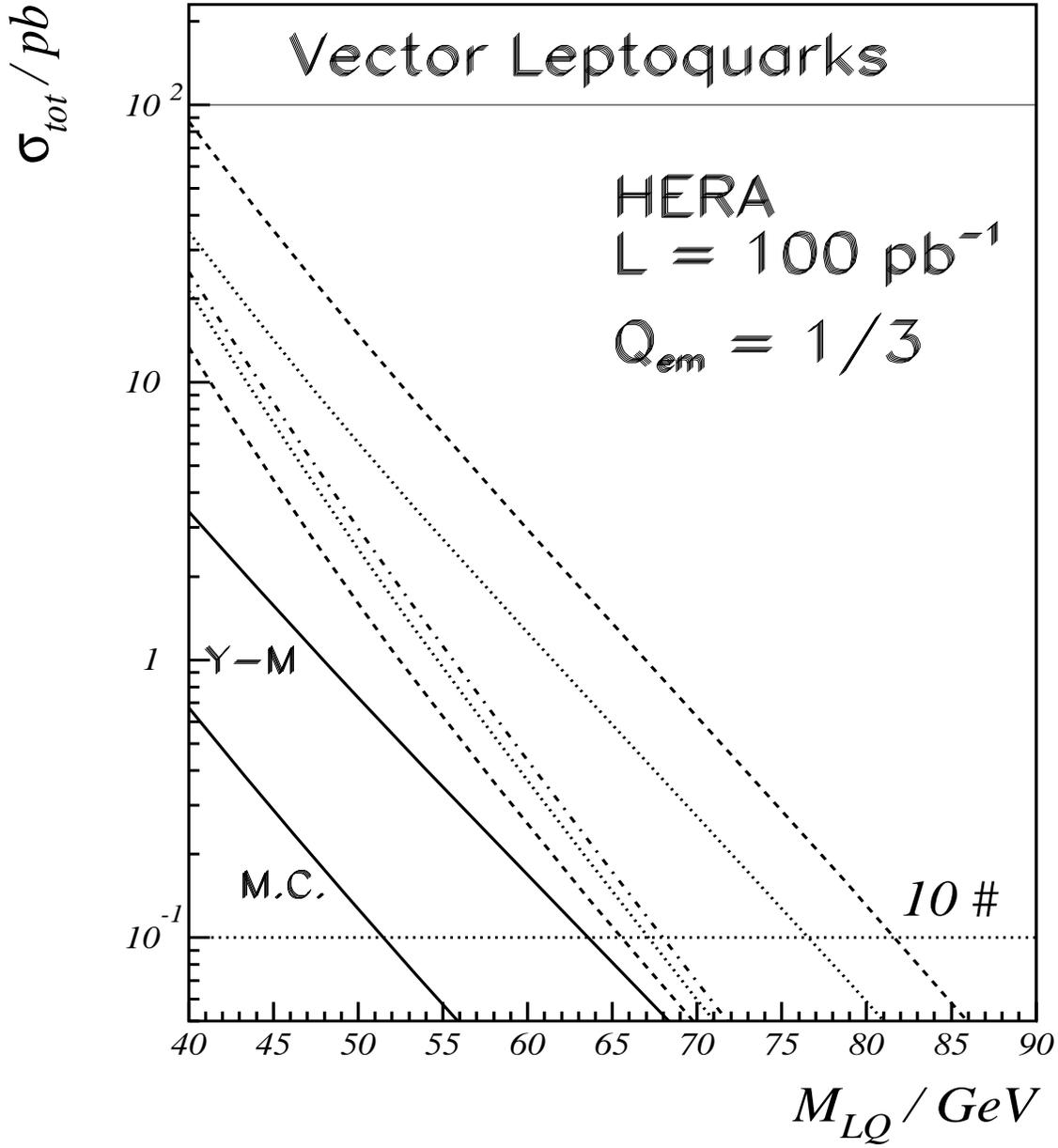

Figure 3a: Integrated cross sections for vector leptoquark pair production for $\sqrt{S} = 314\,\text{GeV}$ and different values of $\kappa_{A,G}$ and $\lambda_{A,G}$. Full lines: minimal coupling (M.C.) $\kappa_{A,G} = 1$, $\lambda_{A,G} = 0$, and Yang–Mills coupling (Y–M) $\kappa_{A,G} = \lambda_{A,G} = 0$. Upper dashed line: $\kappa_{A,G} = \lambda_{A,G} = -1$, lower dashed line: $\kappa_{A,G} = \lambda_{A,G} = 1$; upper dotted line: $\kappa_{A,G} = -1, \lambda_{A,G} = 1$; lower dotted line: $\kappa_{A,G} = 1, \lambda_{A,G} = -1$; dash–dotted line: $\kappa_A = 1, \kappa_G = -1, \lambda_A = 1, \lambda_G = -1$.

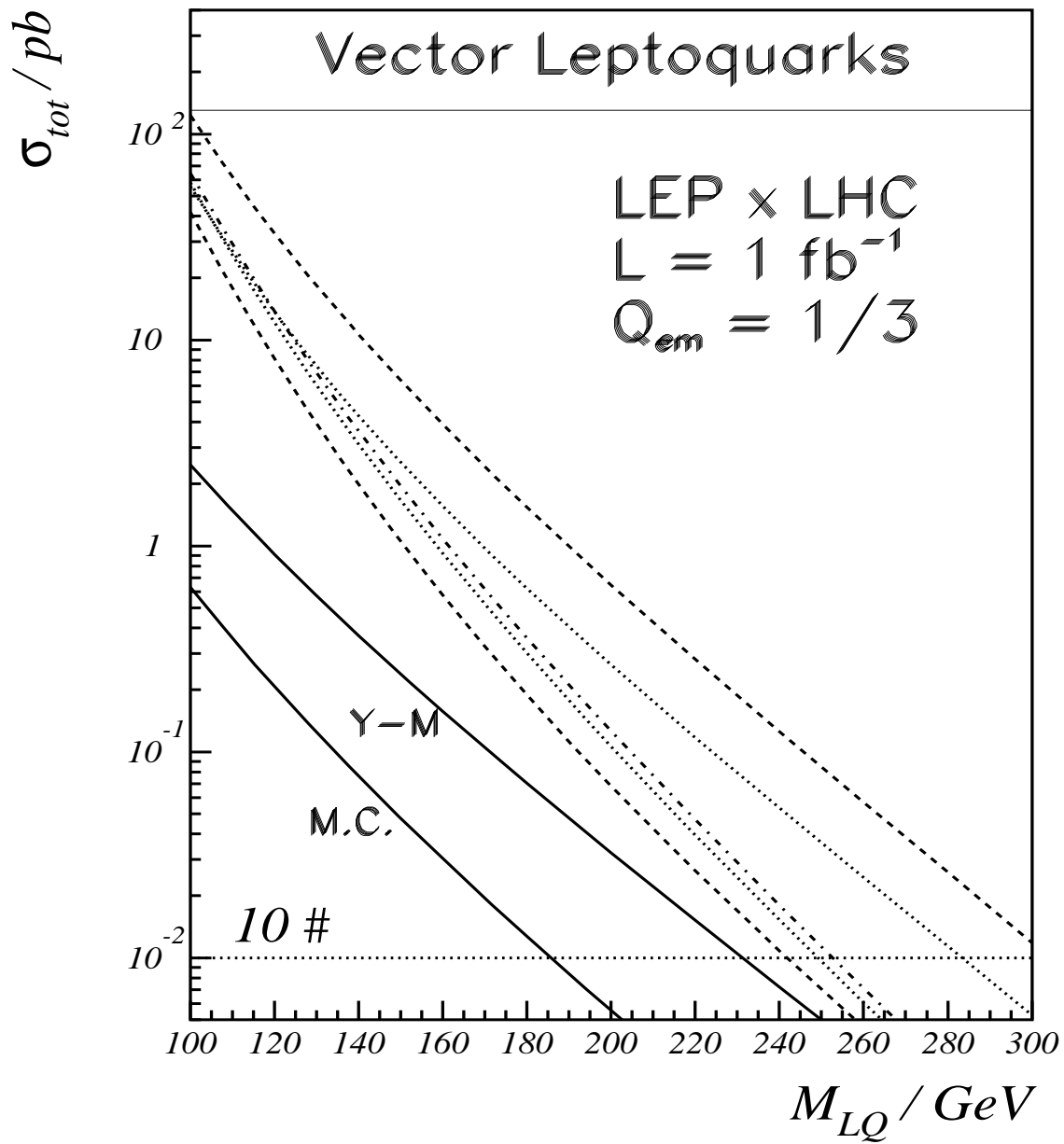

Figure 3b: Integrated cross sections for vector leptoquark pair production for $\sqrt{S} = 1260$ GeV. The other parameters are the same as in figure 3a.